# Vector Laplacian in Spherical Coordinates:
# An Unnoticed Typo in Landau and Lifshitz's Fluid Mechanics Course

Peter Lebedev-Stepanov

Shubnikov Institute of Crystallography, Kurchatov Complex of Crystallography and Photonics, Leninskii prospekt 59, Moscow 119333, Russia

E-mail: lebstep.p@crys.ras.ru

*A previously unaccounted fundamental typo has been discovered in Course of Theoretical Physics (vol. 6) by Landau & Lifshitz; Fluid Mechanics; 1987, Pergamon, page 49, Eqs. (15.21), which corresponds to the same typo in the Russian original of this book. This concerns the first (upper) of the Navier–Stokes equations in spherical coordinates, which includes the r-component of the vector Laplacian, namely, an extra square of the sine in the denominator. This error migrates to secondary publications and can complicate theoretical studies related to the application of the Navier–Stokes equations, which the author of this note has encountered first-hand. This typo is not present in "An Introduction to Fluid Dynamics" by Batchelor (Cambridge. 2000), page 601.*

*The present short article makes a detailed derivation of the r-component of the vector Laplacian from general principles to show what the corresponding Navier–Stokes equation should look like, and that the typo is indeed present. In addition, more compact equivalent forms of the Navier-Stokes equations in spherical coordinates are proposed.*

**1. Statement of the problem**

Fig. 1 and Fig. 2 show screenshots from two authoritative courses on fluid mechanics, presenting the Navier–Stokes equations for an incompressible fluid in a spherical coordinate system. We see that they differ by factor 2 (square of the $\sin\theta$ in the denominator), indicated by the red arrow in Fig. 1. This indicator is absent, for example, in the course [3], as well as in the monograph [4]. The fact that the r-component of the vector Laplacian must have a form corresponding to Fig. 2 also follows from the monograph [5], p. 192.

However, the authority of the course [1] forces us to derive an expression for the r-component of the vector Laplacian in spherical coordinates from first principles to verify that the square in the denominator of the sine in Fig. 1 is indeed a typo requiring correction.

$$\frac{\partial v_r}{\partial t} + (\mathbf{v}\cdot\mathbf{grad})v_r - \frac{v_\theta^2 + v_\phi^2}{r}$$
$$= -\frac{1}{\rho}\frac{\partial p}{\partial r} + \nu\left[\Delta v_r - \frac{2}{r^2\sin^2\theta}\frac{\partial(v_\theta\sin\theta)}{\partial\theta} - \frac{2}{r^2\sin\theta}\frac{\partial v_\phi}{\partial\phi} - \frac{2v_r}{r^2}\right],$$

$$\frac{\partial v_\theta}{\partial t} + (\mathbf{v}\cdot\mathbf{grad})v_\theta + \frac{v_r v_\theta}{r} - \frac{v_\phi^2\cot\theta}{r}$$
$$= -\frac{1}{\rho r}\frac{\partial p}{\partial\theta} + \nu\left[\Delta v_\theta - \frac{2\cos\theta}{r^2\sin^2\theta}\frac{\partial v_\phi}{\partial\phi} + \frac{2}{r^2}\frac{\partial v_r}{\partial\theta} - \frac{v_\theta}{r^2\sin^2\theta}\right],$$

$$\frac{\partial v_\phi}{\partial t} + (\mathbf{v}\cdot\mathbf{grad})v_\phi + \frac{v_r v_\phi}{r} + \frac{v_\theta v_\phi\cot\theta}{r}$$
$$= -\frac{1}{\rho r\sin\theta}\frac{\partial p}{\partial\phi} + \nu\left[\Delta v_\phi + \frac{2}{r^2\sin\theta}\frac{\partial v_r}{\partial\phi} + \frac{2\cos\theta}{r^2\sin^2\theta}\frac{\partial v_\theta}{\partial\phi} - \frac{v_\phi}{r^2\sin^2\theta}\right].$$

Fig.1. Screenshot of the Navier–Stokes equations from Landau & Lifshitz; Fluid Mechanics; 1987, Pergamon, page 49 [1] and its Russian original [2] p. 77. The arrow indicates the erroneously placed square at $\sin\theta$ in the denominator in the upper equation.

$$\frac{\partial u_r}{\partial t} + \mathbf{u}\cdot\nabla u_r - \frac{u_\theta^2}{r} - \frac{u_\phi^2}{r} = -\frac{1}{\rho}\frac{\partial p}{\partial r}$$
$$+\nu\left\{\nabla^2 u_r - \frac{2u_r}{r^2} - \frac{2}{r^2\sin\theta}\frac{\partial(u_\theta\sin\theta)}{\partial\theta} - \frac{2}{r^2\sin\theta}\frac{\partial u_\phi}{\partial\phi}\right\},$$

$$\frac{\partial u_\theta}{\partial t} + \mathbf{u}\cdot\nabla u_\theta + \frac{u_r u_\theta}{r} - \frac{u_\phi^2\cot\theta}{r} = -\frac{1}{\rho r}\frac{\partial p}{\partial\theta}$$
$$+\nu\left\{\nabla^2 u_\theta + \frac{2}{r^2}\frac{\partial u_r}{\partial\theta} - \frac{u_\theta}{r^2\sin^2\theta} - \frac{2\cos\theta}{r^2\sin^2\theta}\frac{\partial u_\phi}{\partial\phi}\right\},$$

$$\frac{\partial u_\phi}{\partial t} + \mathbf{u}\cdot\nabla u_\phi + \frac{u_\phi u_r}{r} + \frac{u_\theta u_\phi\cot\theta}{r} = -\frac{1}{\rho r\sin\theta}\frac{\partial p}{\partial\phi}$$
$$+\nu\left\{\nabla^2 u_\phi + \frac{2}{r^2\sin\theta}\frac{\partial u_r}{\partial\phi} + \frac{2\cos\theta}{r^2\sin^2\theta}\frac{\partial u_\theta}{\partial\phi} - \frac{u_\phi}{r^2\sin^2\theta}\right\}.$$

Fig.2. Screenshot of the Navier–Stokes equations from monography by Batchelor, page 601 [3]. There is no indicator 2 for $\sin\theta$.

## 2. Derivation of the r-component of the vector Laplacian in spherical coordinates

The vector Laplacian is defined by the expression [5-7]:

$$\vec{\nabla^2}\mathbf{E} = \nabla(\nabla\cdot\mathbf{E}) - \nabla\times[\nabla\times\mathbf{E}]. \qquad (1)$$

In the general case, being written by components, Eq. (1) has the form [6,7]:

$$\vec{\nabla}^2 \mathbf{E} = \mathbf{e}_{u1} \left\{ g_{11}^{-\frac{1}{2}} \frac{\partial \Upsilon}{\partial u_1} + \left( \frac{g_{11}}{g} \right)^{\frac{1}{2}} \left[ \frac{\partial \Gamma_2}{\partial u_3} - \frac{\partial \Gamma_3}{\partial u_2} \right] \right\} + \mathbf{e}_{u2} \left\{ g_{22}^{-\frac{1}{2}} \frac{\partial \Upsilon}{\partial u_2} + \left( \frac{g_{22}}{g} \right)^{\frac{1}{2}} \left[ \frac{\partial \Gamma_3}{\partial u_1} - \frac{\partial \Gamma_1}{\partial u_3} \right] \right\} +$$
$$+ \mathbf{e}_{u3} \left\{ g_{33}^{-\frac{1}{2}} \frac{\partial \Upsilon}{\partial u_3} + \left( \frac{g_{33}}{g} \right)^{\frac{1}{2}} \left[ \frac{\partial \Gamma_1}{\partial u_2} - \frac{\partial \Gamma_2}{\partial u_1} \right] \right\}, \tag{2}$$

where the metric coefficients are defined by

$$g_{ii} = \left( \frac{\partial x_1}{\partial u_i} \right)^2 + \left( \frac{\partial x_2}{\partial u_i} \right)^2 + \left( \frac{\partial x_3}{\partial u_i} \right)^2, \quad g = g_{11} g_{22} g_{33}, \tag{3}$$

$x_1$, $x_2$, $x_3$ are rectangular Cartesian coordinates, so that

$$ds^2 = dx_1^2 + dx_2^2 + dx_3^2 = g_{11} du_1^2 + g_{22} du_2^2 + g_{33} du_3^2, \tag{4}$$

$$\Upsilon = g^{-\frac{1}{2}} \left\{ \frac{\partial}{\partial u_1} \left[ \left( \frac{g}{g_{11}} \right)^{\frac{1}{2}} E_1 \right] + \frac{\partial}{\partial u_2} \left[ \left( \frac{g}{g_{22}} \right)^{\frac{1}{2}} E_2 \right] + \frac{\partial}{\partial u_3} \left[ \left( \frac{g}{g_{33}} \right)^{\frac{1}{2}} E_3 \right] \right\}, \tag{5}$$

$$\Gamma_1 = \frac{g_{11}}{g^{\frac{1}{2}}} \left\{ \frac{\partial}{\partial u_2} \left[ g_{33}^{\frac{1}{2}} E_3 \right] - \frac{\partial}{\partial u_3} \left[ g_{22}^{\frac{1}{2}} E_2 \right] \right\}, \tag{6}$$

$$\Gamma_2 = \frac{g_{22}}{g^{\frac{1}{2}}} \left\{ \frac{\partial}{\partial u_3} \left[ g_{11}^{\frac{1}{2}} E_1 \right] - \frac{\partial}{\partial u_1} \left[ g_{33}^{\frac{1}{2}} E_3 \right] \right\}, \tag{7}$$

$$\Gamma_3 = \frac{g_{33}}{g^{\frac{1}{2}}} \left\{ \frac{\partial}{\partial u_1} \left[ g_{22}^{\frac{1}{2}} E_2 \right] - \frac{\partial}{\partial u_2} \left[ g_{11}^{\frac{1}{2}} E_1 \right] \right\}. \tag{8}$$

The structure of Eq. (2) obviously corresponds to the structure of Eq. (1), and:

$$\nabla (\nabla \cdot \mathbf{E}) = g_{11}^{-\frac{1}{2}} \mathbf{e}_{u1} \frac{\partial \Upsilon}{\partial u_1} + g_{22}^{-\frac{1}{2}} \mathbf{e}_{u2} \frac{\partial \Upsilon}{\partial u_2} + g_{33}^{-\frac{1}{2}} \mathbf{e}_{u3} \frac{\partial \Upsilon}{\partial u_3}, \tag{9}$$

$$\nabla \times [\nabla \times \mathbf{E}] = \mathbf{e}_{u1} \left( \frac{g_{11}}{g} \right)^{\frac{1}{2}} \left[ \frac{\partial \Gamma_3}{\partial u_2} - \frac{\partial \Gamma_2}{\partial u_3} \right] + \mathbf{e}_{u2} \left( \frac{g_{22}}{g} \right)^{\frac{1}{2}} \left[ \frac{\partial \Gamma_1}{\partial u_3} - \frac{\partial \Gamma_3}{\partial u_1} \right] + \mathbf{e}_{u3} \left( \frac{g_{33}}{g} \right)^{\frac{1}{2}} \left[ \frac{\partial \Gamma_2}{\partial u_1} - \frac{\partial \Gamma_1}{\partial u_2} \right]. \tag{10}$$

In the spherical system $(r, \theta, \varphi)$, the coordinates are:

$$u_1 = r, \quad 0 \le r < \infty, \tag{11}$$
$$u_2 = \theta, \quad 0 \le \theta < \pi, \tag{12}$$
$$u_3 = \varphi, \quad 0 \le \varphi < 2\pi, \tag{13}$$

$$x_1 = r \sin \theta \cos \varphi, \tag{14}$$
$$x_2 = r \sin \theta \sin \varphi, \tag{15}$$
$$x_3 = r \cos \theta. \tag{16}$$

Metric coefficients are

$$g_{11} = 1, \; g_{22} = r^2, \; g_{33} = r^2 \sin^2 \theta, \; g^{\frac{1}{2}} = r^2 \sin \theta. \tag{17}$$

According to the problem set above, the $u_1$-projection of the vector Laplacian in spherical coordinates is subject to verification:

$$\left(\vec{\nabla}^2 \mathbf{E}\right)_{u_1} = g_{11}^{-\frac{1}{2}} \frac{\partial \Upsilon}{\partial u_1} + \left(\frac{g_{11}}{g}\right)^{\frac{1}{2}} \left[\frac{\partial \Gamma_2}{\partial u_3} - \frac{\partial \Gamma_3}{\partial u_2}\right] = \frac{\partial \Upsilon}{\partial r} + \frac{1}{r^2 \sin\theta}\left[\frac{\partial \Gamma_\theta}{\partial \varphi} - \frac{\partial \Gamma_\varphi}{\partial \theta}\right]. \quad (18)$$

Let us calculate it. To do this, we need to find auxiliary quantities:

$$\Upsilon = \frac{1}{r^2 \sin\theta}\left\{\frac{\partial}{\partial r}\left[r^2 \sin\theta E_r\right] + \frac{\partial}{\partial \theta}\left[\frac{r^2 \sin\theta}{r} E_\theta\right] + \frac{\partial}{\partial \varphi}\left[\frac{r^2 \sin\theta}{r \sin\theta} E_\varphi\right]\right\} = $$
$$= \frac{1}{r^2}\frac{\partial}{\partial r}\left[r^2 E_r\right] + \frac{1}{r \sin\theta}\frac{\partial}{\partial \theta}\left[\sin\theta E_\theta\right] + \frac{1}{r \sin\theta}\frac{\partial E_\varphi}{\partial \varphi} = \nabla \cdot \mathbf{E}, \quad (19)$$

$$\Gamma_\theta = \frac{r^2}{r^2 \sin\theta}\left\{\frac{\partial E_r}{\partial \varphi} - \frac{\partial}{\partial r}\left[r \sin\theta E_\varphi\right]\right\} = \frac{1}{\sin\theta}\frac{\partial E_r}{\partial \varphi} - \frac{\partial}{\partial r}\left[r E_\varphi\right], \quad (20)$$

$$\Gamma_\varphi = \frac{r^2 \sin^2\theta}{r^2 \sin\theta}\left\{\frac{\partial}{\partial r}[rE_\theta] - \frac{\partial E_r}{\partial \theta}\right\} = \sin\theta\left\{\frac{\partial}{\partial r}[rE_\theta] - \frac{\partial E_r}{\partial \theta}\right\}. \quad (21)$$

Next, we find

$$\frac{\partial \Upsilon}{\partial r} = \frac{\partial}{\partial r}\frac{1}{r^2}\frac{\partial}{\partial r}\left[r^2 E_r\right] + \frac{\partial}{\partial r}\frac{1}{r \sin\theta}\frac{\partial}{\partial \theta}[\sin\theta E_\theta] + \frac{\partial}{\partial r}\frac{1}{r \sin\theta}\frac{\partial E_\varphi}{\partial \varphi} = $$
$$= \frac{1}{r^2}\frac{\partial}{\partial r}\left[2rE_r + r^2 \frac{\partial}{\partial r}E_r\right] - \frac{1}{r^2 \sin\theta}\frac{\partial}{\partial \theta}[\sin\theta E_\theta] - \frac{1}{r^2 \sin\theta}\frac{\partial E_\varphi}{\partial \varphi} - $$
$$- \frac{2}{r^3}\left[2rE_r + r^2 \frac{\partial}{\partial r}E_r\right] + \frac{1}{r \sin\theta}\frac{\partial}{\partial \theta}\left[\sin\theta \frac{\partial}{\partial r}E_\theta\right] + \frac{1}{r \sin\theta}\frac{\partial^2 E_\varphi}{\partial \varphi \partial r} \quad (22)$$

or, expanding the derivatives with respect to r in square brackets, we obtain

$$\frac{\partial \Upsilon}{\partial r} = \frac{2E_r}{r^2} + \frac{2}{r}\frac{\partial}{\partial r}E_r + \frac{1}{r^2}\frac{\partial}{\partial r}\left[r^2 \frac{\partial}{\partial r}E_r\right] - \frac{1}{r^2 \sin\theta}\frac{\partial}{\partial \theta}[\sin\theta E_\theta] - \frac{1}{r^2 \sin\theta}\frac{\partial E_\varphi}{\partial \varphi} - $$
$$- \frac{4E_r}{r^2} - \frac{2}{r}\frac{\partial}{\partial r}E_r + \frac{1}{r \sin\theta}\frac{\partial}{\partial \theta}\left[\sin\theta \frac{\partial}{\partial r}E_\theta\right] + \frac{1}{r \sin\theta}\frac{\partial^2 E_\varphi}{\partial \varphi \partial r} = $$
$$= \frac{1}{r^2}\frac{\partial}{\partial r}\left[r^2 \frac{\partial}{\partial r}E_r\right] - \frac{1}{r^2 \sin\theta}\frac{\partial}{\partial \theta}[\sin\theta E_\theta] - \frac{1}{r^2 \sin\theta}\frac{\partial E_\varphi}{\partial \varphi} - $$
$$- \frac{2E_r}{r^2} + \frac{1}{r \sin\theta}\frac{\partial}{\partial \theta}\left[\sin\theta \frac{\partial}{\partial r}E_\theta\right] + \frac{1}{r \sin\theta}\frac{\partial^2 E_\varphi}{\partial \varphi \partial r}. \quad (23)$$

Next, we find:

$$\frac{1}{r^2 \sin\theta}\frac{\partial \Gamma_\theta}{\partial \varphi} = \frac{1}{r^2 \sin\theta}\left(\frac{1}{\sin\theta}\frac{\partial^2 E_r}{\partial \varphi^2} - \frac{\partial}{\partial r}\left[r \frac{\partial E_\varphi}{\partial \varphi}\right]\right) = \frac{1}{r^2 \sin^2\theta}\frac{\partial^2 E_r}{\partial \varphi^2} - \frac{1}{r^2 \sin\theta}\frac{\partial}{\partial r}\left[r \frac{\partial E_\varphi}{\partial \varphi}\right], \quad (24)$$

$$\frac{1}{r^2 \sin\theta} \frac{\partial \Gamma_\varphi}{\partial \theta} = \frac{1}{r^2 \sin\theta} \frac{\partial}{\partial \theta}\left(\sin\theta \frac{\partial}{\partial r}[rE_\theta]\right) - \frac{1}{r^2 \sin\theta} \frac{\partial}{\partial \theta}\left(\sin\theta \frac{\partial E_r}{\partial \theta}\right) =$$
$$= \frac{1}{r^2} \frac{\partial}{\partial r}\left[r \frac{\partial}{\partial \theta} E_\theta\right] + \frac{\cos\theta}{r^2 \sin\theta} \frac{\partial}{\partial r}[rE_\theta] - \frac{1}{r^2 \sin\theta} \frac{\partial}{\partial \theta}\left(\sin\theta \frac{\partial E_r}{\partial \theta}\right) \quad (25)$$

Then

$$\frac{1}{r^2 \sin\theta}\left[\frac{\partial \Gamma_\theta}{\partial \varphi} - \frac{\partial \Gamma_\varphi}{\partial \theta}\right] = \frac{1}{r^2 \sin^2\theta} \frac{\partial^2 E_r}{\partial \varphi^2} + \frac{1}{r^2 \sin\theta} \frac{\partial}{\partial \theta}\left(\sin\theta \frac{\partial E_r}{\partial \theta}\right) -$$
$$- \frac{1}{r^2 \sin\theta} \frac{\partial}{\partial r}\left[r \frac{\partial E_\varphi}{\partial \varphi}\right] - \frac{1}{r^2} \frac{\partial}{\partial r}\left[r \frac{\partial}{\partial \theta} E_\theta\right] - \frac{\cos\theta}{r^2 \sin\theta} \frac{\partial}{\partial r}[rE_\theta] \quad (26)$$

Adding Eqs. (23) and ((26) in accordance with Eq. (18), we obtain

$$\left(\overrightarrow{\nabla^2}\mathbf{E}\right)_{u_1} = g_{11}^{-\frac{1}{2}} \frac{\partial \Upsilon}{\partial u_1} + \left(\frac{g_{11}}{g}\right)^{\frac{1}{2}}\left[\frac{\partial \Gamma_2}{\partial u_3} - \frac{\partial \Gamma_3}{\partial u_2}\right] = \frac{\partial \Upsilon}{\partial r} + \frac{1}{r^2 \sin\theta}\left[\frac{\partial \Gamma_\theta}{\partial \varphi} - \frac{\partial \Gamma_\varphi}{\partial \theta}\right] =$$
$$= \nabla^2 E_r - \frac{2E_r}{r^2} - \frac{1}{r^2 \sin\theta} \frac{\partial}{\partial r}\left[r \frac{\partial E_\varphi}{\partial \varphi}\right] - \frac{1}{r^2} \frac{\partial}{\partial r}\left[r \frac{\partial}{\partial \theta} E_\theta\right] - \frac{\cos\theta}{r^2 \sin\theta} \frac{\partial}{\partial r}[rE_\theta] -$$
$$- \frac{1}{r^2 \sin\theta} \frac{\partial}{\partial \theta}[\sin\theta E_\theta] - \frac{1}{r^2 \sin\theta} \frac{\partial E_\varphi}{\partial \varphi} +$$
$$+ \frac{1}{r \sin\theta} \frac{\partial}{\partial \theta}\left[\sin\theta \frac{\partial}{\partial r} E_\theta\right] + \frac{1}{r \sin\theta} \frac{\partial^2 E_\varphi}{\partial \varphi \partial r} \quad (27)$$

To simplify expression (27), we expand the derivative:

$$\frac{1}{r^2 \sin\theta} \frac{\partial}{\partial r}\left[r \frac{\partial E_\varphi}{\partial \varphi}\right] = \frac{1}{r^2 \sin\theta}\left[\frac{\partial E_\varphi}{\partial \varphi} + r \frac{\partial^2 E_\varphi}{\partial r \partial \varphi}\right]. \quad (28)$$

Next, we expand the derivatives with respect to θ

$$\frac{1}{r \sin\theta} \frac{\partial}{\partial \theta}\left[\sin\theta \frac{\partial}{\partial r} E_\theta\right] = \frac{1}{r} \frac{\partial}{\partial \theta}\left[\frac{\partial}{\partial r} E_\theta\right] + \frac{\cos\theta}{r \sin\theta} \frac{\partial}{\partial r} E_\theta, \quad (29)$$

$$\frac{\cos\theta}{r^2 \sin\theta} \frac{\partial}{\partial r}[rE_\theta] = \frac{\cos\theta}{r^2 \sin\theta} E_\theta + \frac{\cos\theta}{r \sin\theta} \frac{\partial E_\theta}{\partial r}, \quad (30)$$

$$\frac{1}{r^2} \frac{\partial}{\partial r}\left[r \frac{\partial}{\partial \theta} E_\theta\right] = \frac{1}{r} \frac{\partial^2}{\partial \theta \partial r} E_\theta + \frac{1}{r^2} \frac{\partial}{\partial \theta} E_\theta, \quad (31)$$

$$\frac{1}{r^2 \sin\theta} \frac{\partial}{\partial \theta}[\sin\theta E_\theta] = \frac{1}{r^2} \frac{\partial E_\theta}{\partial \theta} + \frac{\cos\theta}{r^2 \sin\theta} E_\theta. \quad (32)$$

The construction in Eq. (27) containing derivatives with respect to θ can be simplified by substituting Eqs. (29)-(32):

$$-\frac{1}{r^2}\frac{\partial}{\partial r}\left[r\frac{\partial}{\partial \theta}E_\theta\right]-\frac{\cos\theta}{r^2\sin\theta}\frac{\partial}{\partial r}[rE_\theta]-\frac{1}{r^2\sin\theta}\frac{\partial}{\partial \theta}[\sin\theta E_\theta]+$$

$$+\frac{1}{r\sin\theta}\frac{\partial}{\partial \theta}\left[\sin\theta\frac{\partial}{\partial r}E_\theta\right]+\frac{1}{r\sin\theta}\frac{\partial^2 E_\varphi}{\partial \varphi \partial r}=-\frac{2}{r^2\sin\theta}\frac{\partial}{\partial \theta}[\sin\theta E_\theta]= \quad (33)$$

$$=-\frac{2}{r^2}\frac{\partial E_\theta}{\partial \theta}-\frac{2\cot\theta}{r^2}E_\theta.$$

Substituting Eqs. (28) and (33) into Eq.(27), we obtain

$$\left(\overrightarrow{\nabla^2}\mathbf{E}\right)_{u_1}=g_{11}^{-\frac{1}{2}}\frac{\partial \Upsilon}{\partial u_1}+\left(\frac{g_{11}}{g}\right)^{\frac{1}{2}}\left[\frac{\partial \Gamma_2}{\partial u_3}-\frac{\partial \Gamma_3}{\partial u_2}\right]=\frac{\partial \Upsilon}{\partial r}+\frac{1}{r^2\sin\theta}\left[\frac{\partial \Gamma_\theta}{\partial \varphi}-\frac{\partial \Gamma_\varphi}{\partial \theta}\right]= \quad (34)$$

$$=\nabla^2 E_r-\frac{2E_r}{r^2}-\frac{2}{r^2\sin\theta}\frac{\partial E_\varphi}{\partial \varphi}-\frac{2}{r^2\sin\theta}\frac{\partial}{\partial \theta}[\sin\theta E_\theta]$$

Here the usual notation for the scalar Laplacian is introduced

$$\nabla^2 E_r=\frac{1}{r^2}\frac{\partial}{\partial r}\left[r^2\frac{\partial}{\partial r}E_r\right]+\frac{1}{r^2\sin\theta}\frac{\partial}{\partial \theta}\left(\sin\theta\frac{\partial E_r}{\partial \theta}\right)+\frac{1}{r^2\sin^2\theta}\frac{\partial^2 E_r}{\partial \varphi^2}. \quad (35)$$

### 3. Conclusion

It is shown that the *r*-component of the vector Laplacian in spherical coordinates is written as

$$\left(\overrightarrow{\nabla^2}\mathbf{E}\right)_{u_1}=\nabla^2 E_r-\frac{2E_r}{r^2}-\frac{2}{r^2\sin\theta}\frac{\partial E_\varphi}{\partial \varphi}-\frac{2}{r^2\sin\theta}\frac{\partial}{\partial \theta}[\sin\theta E_\theta], \quad (36)$$

There is no factor 2 for sine. Therefore, in the Course by Landau and Lifshitz [1] on page 49 and in the Russian-language version of this course [2] on page 77, a typo was made in the first equation (15.21). On the contrary, the monograph [3] is an example where such an error was not made.

Note that sometimes the representation of (35) is used in the form

$$\left(\overrightarrow{\nabla^2}\mathbf{E}\right)_{u_1}=\nabla^2 E_r-\frac{2E_r}{r^2}-\frac{2}{r^2\sin\theta}\frac{\partial E_\varphi}{\partial \varphi}-\frac{2}{r^2}\frac{\partial E_\theta}{\partial \theta}-\frac{2\cot\theta}{r^2}E_\theta, \quad (37)$$

i.e. with the last term expanded ([5], p. 192).

Thus, the correct form of the Navier–Stokes equations for an incompressible fluid is shown in Fig. 2, which corresponds to monograph [3]. We will write them out below for convenience

$$\rho\left(\frac{d\mathbf{V}}{dt}\right)_r=-\frac{\partial}{\partial r}p+\eta\left[\nabla^2 V_r-\frac{2V_r}{r^2}-\frac{2}{r^2\sin\theta}\frac{\partial}{\partial \theta}(V_\theta\sin\theta)-\frac{2}{r^2\sin\theta}\frac{\partial}{\partial \varphi}V_\varphi\right], \quad (38)$$

$$\rho\left(\frac{d\mathbf{V}}{dt}\right)_\theta=-\frac{1}{r}\frac{\partial}{\partial \theta}p+\eta\left[\nabla^2 V_\theta-\frac{V_\theta}{r^2\sin^2\theta}+\frac{2}{r^2}\frac{\partial}{\partial \theta}V_r-\frac{2\cos\theta}{r^2\sin^2\theta}\frac{\partial}{\partial \varphi}V_\varphi\right], \quad (39)$$

$$\rho\left(\frac{d\mathbf{V}}{dt}\right)_\varphi=-\frac{1}{r\sin\theta}\frac{\partial}{\partial \varphi}p+\eta\left[\nabla^2 V_\varphi-\frac{V_\varphi}{r^2\sin^2\theta}+\frac{2}{r^2\sin\theta}\frac{\partial}{\partial \varphi}V_r+\frac{2\cos\theta}{r^2\sin^2\theta}\frac{\partial}{\partial \varphi}V_\theta\right]. \quad (40)$$

The continuity equation for an incompressible fluid, taking into account (19), has the form

$$\nabla \cdot \mathbf{V} = \frac{1}{r^2}\frac{\partial}{\partial r}\left[r^2 V_r\right] + \frac{1}{r\sin\theta}\frac{\partial}{\partial \theta}\left[\sin\theta V_\theta\right] + \frac{1}{r\sin\theta}\frac{\partial V_\varphi}{\partial \varphi} = 0. \quad (41)$$

Note that Eq. (38) can be rewritten in a more compact form, expressing the last two terms from Eq. (41) as:

$$\frac{2}{r^2 \sin\theta}\frac{\partial}{\partial \theta}\left[\sin\theta V_\theta\right] + \frac{2}{r^2 \sin\theta}\frac{\partial V_\varphi}{\partial \varphi} = -\frac{2}{r^3}\frac{\partial}{\partial r}\left[r^2 V_r\right]. \quad (42)$$

Then we obtain another version of the component of the Navier–Stokes equation

$$\rho\left(\frac{d\mathbf{V}}{dt}\right)_r = -\frac{\partial}{\partial r}p + \eta\left[\nabla^2 V_r - \frac{2V_r}{r^2} + \frac{2}{r^3}\frac{\partial}{\partial r}\left[r^2 V_r\right]\right]. \quad (43)$$

Taking into account that

$$\frac{2}{r^3}\frac{\partial}{\partial r}\left[r^2 V_r\right] = \frac{4V_r}{r^2} + \frac{2}{r}\frac{\partial V_r}{\partial r} \quad (44)$$

and

$$\frac{V_r}{r^2} + \frac{1}{r}\frac{\partial V_r}{\partial r} = \frac{1}{r^2}\frac{\partial(rV_r)}{\partial r} \quad (45)$$

one can obtain

$$\rho\left(\frac{d\mathbf{V}}{dt}\right)_r = -\frac{\partial}{\partial r}p + \eta\left[\nabla^2 V_r + \frac{2}{r^2}\frac{\partial(rV_r)}{\partial r}\right]. \quad (46)$$

Eq. (46) is equivalent to Eq. (38), but is much more compact.

There is another way to rewrite this expression. Taking into account

$$\frac{1}{r}\frac{\partial V_r}{\partial r} = \frac{\partial}{\partial r}\frac{V_r}{r} + \frac{V_r}{r^2}, \quad (47)$$

let us rewrite (45) as

$$\frac{1}{r^2}\frac{\partial(rV_r)}{\partial r} = \frac{V_r}{r^2} + \frac{1}{r}\frac{\partial V_r}{\partial r} = \frac{2V_r}{r^2} + \frac{\partial}{\partial r}\frac{V_r}{r}. \quad (48)$$

Substituting Eq. (48) into Eq. (46), one can obtain

$$\rho\left(\frac{d\mathbf{V}}{dt}\right)_r = -\frac{\partial}{\partial r}p + \eta\left[\nabla^2 V_r + \frac{4V_r}{r^2} + \frac{\partial}{\partial r}\frac{2V_r}{r}\right]. \quad (49)$$

or

$$\rho\left(\frac{d\mathbf{V}}{dt}\right)_r = \frac{\partial}{\partial r}\left(\frac{2\eta V_r}{r} - p\right) + \eta\left[\nabla^2 V_r + \frac{4V_r}{r^2}\right]. \quad (50)$$

Note that the other two Navier-Stokes equations (39)-(40) can be rewritten in a similar form, taking into account the definition of the gradient operator in spherical coordinates

$$\rho\left(\frac{d\mathbf{V}}{dt}\right)_\theta = \frac{1}{r}\frac{\partial}{\partial\theta}\left(\frac{2\eta V_r}{r} - p\right) + \eta\left[\nabla^2 V_\theta - \frac{V_\theta}{r^2\sin^2\theta} - \frac{2\cos\theta}{r^2\sin^2\theta}\frac{\partial}{\partial\varphi}V_\varphi\right], \quad (51)$$

$$\rho\left(\frac{d\mathbf{V}}{dt}\right)_\varphi = \frac{1}{r\sin\theta}\frac{\partial}{\partial\varphi}\left(\frac{2\eta V_r}{r} - p\right) + \eta\left[\nabla^2 V_\varphi - \frac{V_\varphi}{r^2\sin^2\theta} + \frac{2\cos\theta}{r^2\sin^2\theta}\frac{\partial}{\partial\varphi}V_\theta\right]. \quad (52)$$

Thus, substituting the gradient operator, the Navier-Stokes equations (38)-(40) can be rewritten in the form

$$\rho\left(\frac{d\mathbf{V}}{dt}\right)_r = \nabla_r\left(\frac{2\eta V_r}{r} - p\right) + \eta\left[\nabla^2 V_r + \frac{4V_r}{r^2}\right]. \quad (53)$$

$$\rho\left(\frac{d\mathbf{V}}{dt}\right)_\theta = \nabla_\theta\left(\frac{2\eta V_r}{r} - p\right) + \eta\left[\nabla^2 V_\theta - \frac{V_\theta}{r^2\sin^2\theta} - \frac{2\cos\theta}{r^2\sin^2\theta}\frac{\partial}{\partial\varphi}V_\varphi\right], \quad (54)$$

$$\rho\left(\frac{d\mathbf{V}}{dt}\right)_\varphi = \nabla_\varphi\left(\frac{2\eta V_r}{r} - p\right) + \eta\left[\nabla^2 V_\varphi - \frac{V_\varphi}{r^2\sin^2\theta} + \frac{2\cos\theta}{r^2\sin^2\theta}\frac{\partial}{\partial\varphi}V_\theta\right]. \quad (55)$$

**Author ORCID**

Peter Lebedev-Stepanov  https://orcid.org/0000-0002-7009-4319